\documentclass[twocolumn,aps,preprintnumbers]{revtex4}
\usepackage{bm}
\usepackage{amsmath}

\renewcommand\>{\rangle}
\renewcommand\d{\partial}
\newcommand\Leff{\mathcal{L}_{\rm eff}}
\newcommand\epsilongs{\epsilon_{\rm gs}}

\newcommand{\npb}[3]{Nucl. Phys. {\bf B#1}, #3 (#2)}

\newcommand{\jetp}[3]{J. Exp. Theor. Phys. {\bf #1}, #3 (#2)}

\begin{document}

\affiliation{Institute for Nuclear Theory, University of Washington,
Seattle WA 98195}

\title{Low-Energy Quantum Effective Action for Relativistic Superfluids}

\author{D.~T.~Son}
%\email{son@phys.washington.edu}
\affiliation{Institute for Nuclear Theory, University of Washington,
Seattle WA 98195}
\preprint{INT-PUB 02-35}

\date{April 2001}

\begin{abstract}
We consider relativistic superfluids where the $U(1)$ baryon symmetry
is spontaneously broken.  We show that all terms in the expansion of
the quantum effective action for the Goldstone field can can be found
to leading order in derivatives, once the equation of state is given.
This enables one to find the all scattering amplitudes between
Goldstone bosons at low energies.  We apply this general result to
quark matter at asymptotically high densities and derive the
low-energy effective action to leading order in $\alpha_s$, and show that
terms containing the fifth and higher powers of the Goldstone field
appear only at the $\alpha_s^2$ level with known coefficients.
\end{abstract}
%\pacs{xxx}
\maketitle

%\section{Introduction}
%\label{sec:intro}
{\em Introduction.}---% 
The behavior of matter at high density is being actively explored in
connection with the physics of neutron stars.  One of the most
important issues is superfluidity, which is predicted for both neutron
liquid~\cite{neutron-super} and quark matter at densities sufficiently
high so that $u$, $d$ and $s$ quarks are present at approximately
equal numbers.  The latter phase, called that color-flavor locked
(CFL) phase~\cite{CFL}, and the closely related ones with kaon or pion
condensation, are the subjects of numerous recent
investigations~\cite{RajagopalWilczek,BedaqueSchaefer,Sonkaon,%
KaplanReddy,KryjevskiNorsen}.
%  The spectrum of this phase consists of
%a massless superfluid Goldstone bosons, a nonet of relatively light
%pseudo-Goldstone bosons, and fermions with much larger gaps.

At length scales relevant for astrophysics, the most important modes
in the superfluids (neutron or quark) are the massless Goldstone
bosons and the superfluid vortices.  Naturally, one would like to be
have an effective theory which contains only these degrees of freedom.
It is in principle possible, at least for quark matter at very high
densities, to derive such a theory from the microscopic theory (i.e.,
QCD): the decay constants and velocities of the Goldstone bosons has
been computed in Ref.~\cite{inverse,BBS,RWZ} and the same techniques
should be extendable to nonlinear terms in the Lagrangian as well.
However, the calculations are technically complex and becomes
impossible at strong coupling.

In this paper, we show that from the equation of state alone one can
construct a fully nonlinear quantum effective action that completely
describes the low-energy dynamics of the superfluid Goldstone field.
The result is given by Eq.~(\ref{final}); this effective action allow
one to find, in particular, the scattering amplitude of an arbitrary
number of Goldstone bosons to leading order in energy.

%\section{Quantum effective action}
%\label{sec:qea}

{\em Quantum effective action.}---% 
The quantum effective action is an useful device to treat the problem
of spontaneous symmetry breaking~\cite{field-theory}.  Consider an
arbitrary field theory in which the ground state breaks a $U(1)$
symmetry spontaneously, and let $\Phi(x)$ be an order parameter.  We
do not require $\Phi$ to be an elementary field; for example, in the
CFL phase the superfluid order parameter is a colorless six-quark
composite operator.  The nature of $\Phi$ will not play any role in
the subsequent discussion.

The partition function is defined as
\begin{equation}
  Z[J] = \int\!\mathcal{D}\phi_i\,
  \exp\biggl(iS + i\!\int\!d^4x\, J(x)\Phi(x)\biggr) \,,
\end{equation}
where the path integral is taken over all elementary fields $\phi_i$
in the theory.  The partition function is the generating functional
for all Green functions constructed from $\Phi$.  On the other hand,
\begin{equation}
  W[J] = -i \ln Z[J]
\end{equation}
generates the connected Feynman graphs constructed from $\Phi$.  The
quantum effective action $\Gamma[\Phi]$ is defined as the Legendre
transform of $W[J]$,
\begin{equation}
  \Gamma[\Phi] = W[J] - J\Phi\,, \quad 
  \Phi = \frac{\delta W[J]}{\delta J}\,,\quad
  J = \frac{\delta\Gamma[\Phi]}{\delta\Phi} \,.
\end{equation}
In perturbation theory $\Gamma[\Phi]$ is the sum of one-particle
irreducible graphs, with external legs cut off.  In other words,
$\Gamma[\Phi]$ reproduces, at tree level, all Green functions
constructed from $\Phi$.

The configuration where $\Gamma[\Phi]$ attains its minimum is the
ground-state average of $\Phi$ at vanishing source $J=0$.  In the case
of spontaneous symmetry breaking, the minimum is degenerate.  At the
minima $\Gamma[\Phi]$ is equal to $W[0]$, and is proportional to the
ground state energy,
\begin{equation}\label{GammaE0}
  \min_{\{\Phi\}} \Gamma[\Phi] = W[0] = -E_0 T \,,
\end{equation}
where $T$ is the total time interval, and $E_0$ is the ground state
energy of the Hamiltonian corresponding to the action $S$.

%\section{QCD at finite baryon density}
%\label{sec:derivation}

{\em Derivation of the effective action for the Goldstone mode.}---%
We shall consider QCD at finite baryon chemical potential $\mu$, which
is described by the Lagrangian
\begin{equation}
  \mathcal{L} = \mathcal{L}_0 + \frac{\mu}3\bar q\gamma^0 q \,,
  \label{L}
\end{equation}
where $\mathcal{L}_0$ is the usual QCD Lagrangian at zero chemical
potential, $q$ is the quark field.  Note that $\mathcal{L}_0$ is
Lorentz invariant, and the last term in Eq.~(\ref{L}) is the only
Lorentz-breaking term in the Lagrangian.  This fact will become useful
in our further discussion.  The Lagrangian~(\ref{L}) possesses a U(1)
symmetry generated by the baryon charge,
\begin{equation}
  q \to e^{i\alpha/3} q \,,
  \label{qtrans}
\end{equation}
which is not broken by the QCD vacuum, but may be broken by the ground
state at finite $\mu$, say, due to color superconductivity~\cite{CFL}.
The order parameter $\Phi\sim qqqqqq$ has baryon number 2.  To be
general, let us assume the baryon charge of the order parameter is
$M$; so $U(1)_B$ is broken to $Z_M$.

The entire formalism of quantum effective action does not assumes
Lorentz invariance at any stage and can be applied to the
theory~(\ref{L}).  Because the Lagrangian contains $\mu$ as an
external parameter, $\Gamma$ is a function of $\mu$ as well as a
functional of $\Phi$,
\begin{equation}
  \Gamma = \Gamma[\mu,\, \Phi] \,.
\end{equation}

For the sake of constraining the form of $\Gamma[\mu,\Phi]$, it is
useful to generalize Eq.~(\ref{L}) to
\begin{equation}\label{LA}
  {\cal L} = {\cal L}_0 + \frac13 A_\mu(x) \bar q\gamma^\mu q \,,
\end{equation}
and treat $A_\mu$ as an arbitrary background field, and put $A_0=\mu$,
$A_i=0$ only at the last stage.  Now $\Gamma=\Gamma[A_\mu,\Phi]$,
which is a functional of both the background $A_\mu$ and $\Phi$.
Since the Lagrangian~(\ref{LA}) possesses a gauge symmetry
\begin{equation}
  q\to q e^{i\alpha/3}\,,\qquad A_\mu\to A_\mu+\d_\mu\alpha\,,
\end{equation}
the quantum effective action also respects this symmetry,
\begin{equation}
  \Gamma[A_\mu,\,\Phi] = \Gamma[A_\mu+\d_\mu\alpha,\, \Phi e^{iM\alpha}] \,.
\end{equation}

Our final goal is to derive the quantum effective action for the
Goldstone field, which is the phase of $\Phi$,
\begin{equation}\label{varphi-def}
  \Phi = |\Phi| e^{iM\varphi}\,,\quad \varphi \sim \varphi + \frac{2\pi}M\,,
\end{equation}
so one should try to integrate out the amplitude $|\Phi|$ and keep
only $\varphi$.  In the effective action formalism, the procedure of
``integrating out'' simply amounts to minimization with respect to
$|\Phi|$, because $\Gamma$ should be used strictly at tree level.
Therefore we define
\begin{equation}\label{min-modulus}
  \Gamma[A_\mu,\,\varphi] \equiv
  \min_{\{|\Phi|\}} \Gamma[A_\mu,\, |\Phi|e^{iM\varphi}]\,.
\end{equation}
The gauge symmetry satisfied by $\Gamma[A_\mu,\,\varphi]$ is
\begin{equation}\label{GammaAphigauge}
  \Gamma[A_\mu,\, \varphi] = \Gamma[A_\mu+\d_\mu\alpha,\, \varphi+\alpha]\,.
\end{equation}
In particular, if $\alpha$ is constant in space-time,
Eq.~(\ref{GammaAphigauge}) implies the invariance of the effective action
with respect to shifting $\varphi$ by any constant value.

Let us now make an expansion of $\Gamma[A_\mu,\,\varphi]$ in power
series of the field $\varphi$ as well as of the spatial derivatives.
Due to the shifting symmetry mentioned above, at any given, say,
$n$-th, power of $\varphi$ the number of spatial derivative is at
least $n$: there should be no $\varphi$ without a spatial derivative
staying in front.  If one keeps, at each power of $\varphi$, only
terms with the smallest number of spatial derivatives (symbolically,
only $\d^n\varphi^n$ terms but not $\d^m\varphi^n$ with $m>n$), then
the effective action density $\Leff$ depends only on $\d_\mu\varphi$,
\begin{equation}\label{Gammaapprox}
  \Gamma[A_\mu,\,\varphi] \approx
% \Gamma[A_\mu,\,\d_\mu\varphi]
  \int\!d^4x\,\mathcal{L}_{\rm eff} (A_\mu,\, \d_\mu\varphi)
  \,.
\end{equation}
This will be the only approximation made to derive the final
result~(\ref{final}).  After making this approximation, we still can
use $\Gamma$ to compute the scattering amplitude between any number of
Goldstone bosons to leading power in external momentum.  For example,
we can compute the leading ${\cal O}(p^n)$ behavior of $n$-point Green
functions of Goldstone bosons, but not the corrections suppressed by
extra powers of momenta.  In this way our effective action $\Gamma$
becomes additionally ``low-energy'' in the sense of, e.g., the chiral
perturbation theory, with the difference that the expansion parameter
in our case is $p\varphi$ (momentum times field) instead of $p$
(momentum), and the effective action is strictly tree-level.  We shall
argue below that Eq.~(\ref{Gammaapprox}) is valid below the gap
energy.

The U(1) gauge symmetry~(\ref{GammaAphigauge}) further restricts the
low-energy effective action to
\begin{equation}
%\begin{split}
  \Gamma[A_\mu,\,\varphi] = 
%\Leff(D_\mu\varphi) \equiv
%  \Gamma[\d_\mu\varphi - A_\mu]
  \int\!d^4x\, \Leff(D_\mu\varphi)
  \,,
%\end{split}
\end{equation}
where $D_\mu\varphi\equiv\d_\mu\varphi-A_\mu$.  $\Leff$ now depends on
one variable instead of two.  In principle, $\Leff$ can depend on the
field strenth $F_{\mu\nu}=\d_\mu A_\nu-\d_\nu A_\mu$, but since at the
end we will substitute $A_\mu=(\mu,0)$, we can restrict ourselves to
constant fields $A_\mu$ from now on.

To find the form of $\Leff[D_\mu\varphi]$, we perform a minimization
of $\Gamma$ over $\varphi$. Assuming that the minimum is achieved at
constant fields $\varphi$ independent of $x$, the result becomes
$V_4\Leff[-A_\mu]$ ($V_4$ is the total four-volume).  If we can find
this function, then $\Leff[D_\mu\varphi]$ is obtained by a simple
replacement $-A_\mu\to D_\mu\varphi$.

Because $\Gamma[\varphi]$ was obtained from $\Gamma[\Phi]$ by
minimizing with respect to the modulus $|\Phi|$
[Eq.~(\ref{min-modulus})], by taking the minimization over the phase
$\varphi$ we have performed a full minimization over all
configurations of $\Phi$.  According to Eq.~(\ref{GammaE0}), $\Leff$
must concide, up to a sign, with the ground state energy density
$\epsilongs$ at constant external field $A_\mu$,
\begin{equation}
  \Leff(-A_\mu) = -\epsilongs(A_\mu)\,.
\end{equation}
Moreover, since $A_\mu$ is the only Lorentz breaking term in the
Lagrangian~(\ref{LA}), the result must depend only on $A_\mu A^\mu$.
Therefore, it is sufficient to compute $\epsilongs$ for $A_\mu$
aligned along the time axis.

Let us look at $\epsilongs$ for $A_\mu=(\mu,0)$ from the thermodynamic
point of view.  It is the ground state energy, per unit volume, of the
Hamiltonian
\begin{equation}
H
= H_0 -\mu Q_B\,, 
\end{equation}
where $H_0$ is the vacuum Hamiltonian of QCD and $Q_B$ is the total
baryon number.  We can rewrite that as
$\epsilongs=(\<H_0\>-\mu\<Q_B\>)/V$, where the averaging is taken over
the ground state.  We recognize that $\epsilongs=-P$, where $P$ is the
thermodynamic pressure at chemical potential $\mu$.  The
dependence of $P$ on $\mu$ cannot be further restricted; we shall
assume that the equation of state $P=P(\mu)$ is given and move on.  We
now have
\begin{equation}
  \Leff(-A_\mu) = P\left((A_\mu A^\mu)^{1/2}\right)\,,
\end{equation}
and hence, the full low-energy effective action is completely
determined from the equation of state,
\begin{equation}
  \Gamma[A_\mu,\,\varphi] = 
  \int\!d^4x\, P\left(D_\mu\varphi D^\mu\varphi)^{1/2}\right)\,.
\end{equation}
Putting, in this expression, $A_\mu=(\mu,\,0)$, we obtain finally
\begin{equation}\label{final}
  \Gamma[\mu,\,\varphi] = \int\!d^4x\, 
  P\left(\sqrt{(\d_0\varphi-\mu)^2-(\d_i\varphi)^2}\right)\,.
\end{equation}
Again, the functional dependence of $P$ in Eq.~(\ref{final}) on its
argument is the same as in the equation of state $P=P(\mu)$, which is
an external element in our formalism.  One can expand
Eq.~(\ref{final}) over powers of $\d_0\varphi$ and $(\d_i\varphi)^2$
and find all vertices in the effective actions.

%\section{Example: quark matter at high densities}

{\em Quark matter at high densities.}---%
To illustrate how Eq.~(\ref{final}) can be applied a concrete
situation, let us consider three-flavor quark matter at very high
chemical potential, where the strong coupling $\alpha_s$ is small and
the ground state is the CFL superfluid~\cite{CFL}.  The pressure, 
to leading order in $\alpha_s$, is the same as for a free quark gas,
\begin{equation}\label{eqofstate}
  P(\mu) = \frac{N_c N_f}{12\pi^2} \mu^4\,.
\end{equation}
In the CFL phase $N_c=N_f=3$; we howerver shall keep $N_c$ and $N_f$
explicit.  The effective Lagrangian is now
\begin{eqnarray}
  &&\Leff(\varphi) = \frac{N_c N_f}{12\pi^2}
  \Bigl[(\d_0\varphi-\mu)^2-(\d_i\varphi)^2\Bigr]^2
  \label{Lunexpanded} \\
  &&=\frac{N_cN_f}{12\pi^2} \Bigl[\mu^4-4\mu^3\d_0\varphi
  +6\mu^2(\d_0\varphi)^2-2\mu^2(\d_i\varphi)^2 \nonumber \\
  &&\qquad- 4\mu\d_0\varphi \d_\mu\varphi \d^\mu\varphi
  + (\d_\mu\varphi\d^\mu\varphi)^2\Bigr]\,. \label{Lexpanded}
\end{eqnarray}
It is interesting that in this case the series over $(\d\phi)$
terminates at the four-power terms.  The first term in the
expansion~(\ref{Lexpanded}) is the constant equilibrium pressure; the
second term is a total derivative and can be neglected if one is not
interested in vortices where $\varphi$ is multi-valued
(in which case it leads the Magnus force).  The ratio of the
coefficients of the two quadratic terms in Eq.~(\ref{Lexpanded}) gives
the velocity of the Goldstone boson; as we see it is equal to
$1/\sqrt{3}$, which is the hydrodynamic sound speed in an
ultrarelativistic gas.

Moreover, the relation determining the speed of the Goldstone boson
$u_s$, the pressure $P$ and the energy density $\epsilon$,
\begin{equation}\label{hydrosound}
  u_s^2 = \frac{\d P}{\d\epsilon}\,,
\end{equation}
is exact for any equation of state, not just for $P\sim\mu^4$.  To
show that one expands Eq.~(\ref{final}) to the quadratic order and
finds
\begin{equation}
  u_s^2 = \frac1{\mu} \biggl(\frac {\d P}{\d\mu}\biggr)
  \biggl(\frac{d^2P}{d\mu^2}\biggr)^{-1}
  = \frac {n}{\mu} \frac {d\mu}{d n}\,,
\end{equation}
where $n=dP/d\mu$ is the baryon density.  By using the
zero-temperature thermodynamic relations $dP=nd\mu$,
$d\epsilon=\mu dn$, one arrives to Eq.~(\ref{hydrosound}).  It is
nontrivial that the speed of the Gondstone boson is exactly equal to
that of the hydrodynamic sound, even as we are at zero temperature
outside any hydrodynamic regime.  
%The existence of a mode propagating
%at the hydrodynamic speed is not guaranteed: for example, in the
%presence of two chemical potentials coupled to two broken U(1) charges
%such a mode does not generally exist.

The last two terms in Eqs.~(\ref{Lexpanded}) describe the interaction
between the Goldstone bosons. The knowledge of $\mu$ turns out to be
sufficient to find the scattering amplitude between the Goldstone
bosons.  The Lagriangian~(\ref{Lexpanded}) should be useful for
computing, e.g., the neutrino emission and absorption 
rates~\cite{neutrino} or the
thermal conductivitity~\cite{ShovkovyEllis} of the superfluid core of
neutron stars.

One can easily compute corrections to Eq.~(\ref{Lunexpanded}) by
including corrections to the equation of state.  Because at small
coupling $g$ the superconducting gap is exponentially suppressed as
$e^{-1/g}$~\cite{Songap}, it never appears in the perturbative
expansion of the pressure over $\alpha_s$.  One hence can use
Ref.~\cite{McLerran}, which provides $P$ to the two-loop order.  For
example, with the first correction Eq.~(\ref{eqofstate}) becomes
\begin{equation}\label{Pnext}
  P(\mu) = \frac{N_cN_f}{12\pi^2} \biggl(
  1 - \frac{3(N_c^2-1)}{N_c} \frac{\alpha_s(\mu)}\pi\biggr) \mu^4\,.
\end{equation}
The effective action is obtained by replacing $\mu$ in
Eq.~(\ref{Pnext}) by
$[(\d_\mu\varphi{-}\mu)^2{-}(\d_i\varphi)^2]^{1/2}$.  One immediately
sees that the running of $\alpha_s$ is essential to obtains terms of
different structure than those appearing in Eq.~(\ref{Lexpanded}); in
particular, terms with five and more Goldstone fields appears only in
the $\alpha_s^2$ order with a coefficient proportional to the beta
function coefficient $\beta_0$.

Notice that if instead of $P\sim\mu^4$ the equation of state was
$P\sim\mu^2$, then the superfluid mode would move with the speed of
light and is non-interacting.  Such a situation occurs, for example,
in QCD at finite isospin chemical potential $\mu_I$ when
$m_\pi\ll\mu_I\ll 4\pi f_\pi$~\cite{isospin}.

%\section{Relation to hydrodynamics}

{\em Hydrodynamic interpretation.}---% 
Although the effective action we have just derived is a fully quantum
object, it is instructive to derive from it the classical equation of 
motion.  We find
\begin{equation}\label{eqmotion}
  \d_\mu\biggl(n_0\frac {D_\mu\varphi}{\mu_0} \biggr) = 0\,,
\end{equation}
where the covariant derivatives are taken on the background
$A_\mu=(\mu,0)$, and
$\mu_0\equiv(D_\mu\varphi\,D^\mu\varphi)^{1/2}$ and $n_0\equiv
dP/d\mu|_{\mu=\mu_0}$ were introduced as {\em notations}.  
If we furthermore denote
\begin{equation}
  u_\mu = -\frac{D_\mu\varphi}{\mu_0}\,,
\end{equation}
so that $u^\mu u_\mu=1$, then the equation of motion (\ref{eqmotion})
obtains the form of the hydrodynamic conservation law $\d_\mu(n_0
u^\mu)=0$, where $u^\mu$ plays the role of the fluid velocity and
$n_0$ plays the role of the baryon density in the frame locally
comoving with the fluid.

Furthermore, let us construct the stress-energy tensor from the
action~(\ref{final}).  If we use the standard formula,
\begin{equation}\label{tildeT}
  \tilde T^{\mu\nu} = \d^\nu\varphi\frac{\d\Leff}{\d(\d_\mu\varphi)} -
    g^{\mu\nu}\Leff\,,
\end{equation}
then the conservation law $\d_\mu\tilde T^{\mu\nu}=0$ is satisfied,
but $\tilde T^{\mu\nu}$ is not symmetric.  One can correct the problem
by adding to $\tilde T^{\mu\nu}$ an extra term,
\begin{equation}\label{T}
   T^{\mu\nu} = \tilde T^{\mu\nu} + A^\nu n_0 u^\mu\,,
\end{equation}
which does not spoil the conservation law (since $A^\nu$ is constant
and $\d_\mu(n_0 u^\mu)=0$), but renders $T^{\mu\nu}$ symmetric. 
Substituting Eq.~(\ref{final}) into Eqs.~(\ref{tildeT}) and (\ref{T}),
one finds that the new stress-energy tensor
formally has the same form as that of a perfect fluid,
\begin{equation}\label{fluid}
%\begin{split}
  T^{\mu\nu} = \frac{n_0}{\mu_0} D^\mu\varphi D^\nu\varphi 
  - g^{\mu\nu}\Leff
%\\
  = (\epsilon+P) u^\mu u^\nu - g^{\mu\nu} P\,.
%\end{split}
\end{equation}
In order to perform the last transformation in Eq.~(\ref{fluid}) one
has to recall that $\Leff=P$ and use the thermodynamic relation
$n_0\mu_0=\epsilon+P$, where both $\epsilon$ and $P$ are computed at
the chemical potential $\mu_0$.  However, in contrast to the usual
hydrodynamics where $\mu_0$ and $u^\mu$ are truly independent variables
(except for the condition $u^\mu u_\mu=1$), in our case there is only
one independent dynamical field, which can be chosen to be $\varphi$.

To understand more fully the meaning of $u^\mu$ from the hydrodynamic
point of view, let us go to the norelativistic limit (e.g., consider
neutron fluid).  This limit corresponds to baryon chemical potentials
$\mu$ is close to the nucleon mass $m_N$.  Assuming $\d_\mu\varphi\ll
m_N$, then
\begin{equation}\label{nonrel}
  u^i = \frac1{m_N}\d_i\varphi\,,
\end{equation}
which is the well-known relation between the superfluid velocity and
phase gradient in nonrelativistic superfluids [the mass staying in
Eq.~(\ref{nonrel}) is $m_N$ instead of that for the Cooper pair $2m_N$
because the factor of 2 has been absorbed in the definition of
$\varphi$, Eq.~(\ref{varphi-def})].

The relativistic generalization of the superfluid hydrodynamic
equations has been discussed before~\cite{LebedevKhalatnikov,Carter}.
Up to some field redefinitions, the equations derived in
Refs.~\cite{LebedevKhalatnikov,Carter} (at zero temperature) coincides
with our equation of motion for $\varphi$.  The approach taken in
Refs.~\cite{LebedevKhalatnikov,Carter} is to start from the known
nonrelativistic hydrodynamic equations and generalize them to the
relativistic case.  We have arrived to the same equation following a
different path which emphasizes the role play by the symmetries
(Lorentz and baryon number).  We also achieved an understanding of
nature of the action~(\ref{final}) as the low-energy quantum effective
action for the Goldstone mode, which can be used not only at the
classical level but also for purely quantum calculations, for example,
of the scattering amplitudes between Goldstone bosons.  However, the
formalism described here does not give the hydrodynamic equations at
finite temperature.

{\em Conclusion.}---% 
It is surprising that so much information about the superfluid
Goldstone bosons and their interaction can be infered from the
equation of state alone.  One should, however, excercise some care
with the effective action~(\ref{final}).  In this Lagragian terms with
more derivatives than fields have been omitted, so its applicability
is limited to energies below some scale.  This scale must be of order
of the gap energy $\Delta$, which does not appear in the leading order
effective action~(\ref{final}).  Indeed, explicit
calculations~\cite{Zarembo} show that the dispersion curve for the
Goldstone boson is no longer linear when its energy is comparable to the gap,
which means that terms of the type $p^4\varphi^2$ becomes important at
that scale.  It is natural to expect corrections to terms with higher
powers of $\varphi$ to become important at the same scale.

I am indebted to S.~Beane, D.~B.~Kaplan, Yu.~Kov\-che\-gov,
I.~Shovkovy, A.~Starinets, and M.~Stephanov for stimulating
discussions.  This work is supported, in part, by DOE Grant No.\
DOE-ER-41132 and by the Alfred P.~Sloan Foundation.

\end{document}